\documentclass[12pt,preprint]{aastex}

\newcommand{\eb}{\begin{equation}}
\newcommand{\ee}{\end{equation}}
\newcommand{\ergs}{erg~s$^{-1}$}
\newcommand{\kms}{km~s$^{-1}$}
\newcommand{\masyr}{mas~yr$^{-1}$}
\newcommand{\msun}{$M_{\sun}$}

\shorttitle{Common proper motion companions}
\shortauthors{Makarov et al.}
\begin{document}

\title{Common Proper Motion Companions to Nearby Stars: Ages and Evolution} 
\author{V. V. Makarov}
\affil{Michelson Science Center, California Institute of Technology, 770 S. Wilson Ave.,
MS 100-22, Pasadena, CA 91125}
\author{N. Zacharias \and G.S. Hennessy}
\affil{US Naval Observatory, 3450 Massachusetts Ave, NW, Washington, DC 20392-5420}
\email{vvm@caltech.edu}

\begin{abstract}
A set of 41 nearby stars (closer than 25 pc) is investigated which have very wide binary and common proper motion (CPM) companions at projected separations between $1000$ and $200\,000$ AU.
These companions are identified by astrometric positions and proper motions from the NOMAD catalog.
Based mainly on measures of chromospheric and X-ray activity, age estimation is obtained for most
of 85 identified companions. Color -- absolute magnitude
diagrams are constructed to test if CPM companions are physically related to the primary
nearby stars and have the same age.
Our carefully selected sample includes three remote white dwarf companions to main sequence stars
and two systems (55 Cnc and GJ 777A) of multiple planets and distant stellar companions.
Ten new CPM companions, including three of extreme separations, are found. Multiple hierarchical systems are abundant;
more than 25\% of CPM components are spectroscopic or astrometric binaries or multiples themselves.
Two new astrometric binaries are discovered among nearby CPM companions, GJ 264 and HIP 59000
and preliminary orbital solutions are presented. The Hyades kinematic group (or stream) is presented
broadly in the sample, but we find
few possible thick disk objects and none halo stars. It follows from our investigation that
moderately young (age $\lesssim 1$ Gyr) thin disk dwarfs are the dominating species in the
near CPM systems, in general agreement with the premises of the dynamical
survival paradigm. Some of the multiple stellar systems with remote CPM companions probably undergo the
dynamical evolution on non-coplanar orbits, known as the Kozai cycle.

\end{abstract}

\section{Introduction}
\label{firstpage}
Components of wide stellar binaries and common proper motion pairs have been drawing considerable interest for many years. Despite the increasing accuracy of observations and
the growing range of accessible wavelengths, the origin of very wide, weakly bound or unbound systems
remains an open issue. \citet{lep7} estimated that at least 9.5\% of stars within 100 pc have companions
with projected separations greater than 1000 AU. The renewed interest was boosted by the detection of a dearth of substellar mass companions in spectroscopic binaries, and by the attempts to account for
the missing late-type members of the near solar neighborhood.

The main objective of this paper is to investigate a well-defined set of nearby stars in very
wide CPM pairs and to discover new pairs, possibly with low-mass companions. The secondary goal of our
investigation is to establish or estimate the age and evolutionary status of bona fide companions
using a wide range of available astrometric and astrophysical data. The origin and status of
wide CPM systems is still a mystery, because most of them are likely unbound or very
weakly bound and are expected to be easily disrupted in dynamical interactions with other stars
or molecular clouds (\S~\ref{surv.sec}). The nearest stars to the Sun, $\alpha$ Cen and
Proxima Cen, form a wide pair which may be on a hyperbolic orbit \citep{ano}. It is expected
that such systems should be mostly young, or belong to moving groups, remnant
clusters or associations, but this has not yet been demonstrated on a representative sample. It is
not known if the companions formed together and have the same age. We combine age-related parameters and data,
including color-absolute magnitude diagrams (\S~\ref{hr.sec}), chromospheric activity 
indeces (\S~\ref{chr.sec}), coronal X-ray luminosity (\S~\ref{x.sec}),
multiplicity parameters (\S~\ref{bin.sec}) and kinematics (\S~\ref{skg.sec}) to shed light on this problem.

Previous investigations in the field are too numerous to be listed, but a few papers in considerable overlap with this study should be mentioned. \citet{pov} published a catalog
of 305 nearby wide binary and 29 multiple systems. They discussed the importance of moving
groups for separating different species of wide binaries and tentatively assigned 32 systems to the Hyades stream (called supercluster following Eggen's nomenclature), and 14 to the
Sirius stream. \citet{sal} undertook a comprehensive revision of the Luyten catalog for
approximately 44\% of the sky, drastically improving precision of epoch 2000 positions and proper
motions, and supplying the stars with NIR magnitudes from 2MASS. This allowed \citet{gou} to estimate,
for the first time, trigonometric parallaxes of 424 common proper motion companions to Hipparcos stars
with reliable parallaxes. This extrapolation of parallaxes to CPM companions is justified for high-proper
motion pairs where the rate of chance alignments is small. There is significant overlap between
the sample investigated in this paper and the catalog of \citet{gou}, although we did not use the latter
as a starting point for our selection. We are also employing the parallax extrapolation technique
for dim companions not observed by Hipparcos when constructing color-magnitude diagrams in this paper.

\section{Selecting CPM systems}
\label{sel.sec}
Our selection of candidate CPM systems was based on the
Naval Observatory Merged Astrometric Dataset (NOMAD\footnote{http://nofs.navy.mil/nomad}), 
\citep{znom}, which provides an all-sky catalog of astrometric and photometric
data. NOMAD includes astrometric data from the UNSO-B \citep{mon}, UCAC2
\citep{zac}, Hipparcos (ESA 1997), Tycho-2 (H{\o}g
et al. 2002) catalogs and the ''Yellow Sky" data set (D. Monet, private com.), 
supplemented by $BVR$ optical photometry, mainly from USNO-B, and $JHK$ near-IR photometry from 2MASS.
This catalog covers the entire magnitude range from the
brightest naked eye stars to the limit of the POSS survey plates (about 21st mag).
The largest systematic positional errors are estimated for the Schmidt
plate data used in the USNO-B catalog, with possible local offsets up to about 300 mas.
Systematic errors in UCAC2 and 2MASS are much smaller \citep{zac06}.
Thus, possible systematic errors of proper motions in NOMAD for the entries taken from the USNO-B
catalog can be as large as 10 \masyr, and for faint UCAC2 stars up to about 5 \masyr.
Internal random errors of proper motions are given for all stars in the NOMAD catalog;
these are typically 5 to 10 \masyr\ for faint stars.

We used the following four criteria to select candidate CPM systems for this
investigation: 1) at least one of the components should be listed in both Hipparcos
and NOMAD; 2) the Hipparcos parallax of the primary component should be
statistically reliable and greater than 40 mas (distance less than 25 pc);
3) at least one companion to the primary is found in NOMAD within $1.5\degr$ at
epoch J2000, whose proper motion is within a tolerance limit of the primary's
proper motion; 4) the companion should be clearly visible in both DSS and 2MASS surveys,
and be listed in 2MASS with $J$, $H$ and $K_s$ magnitudes. 
Some 1200 stars from the Hipparcos Catalogue with a parallax
of 40 mas or larger were selected as initial targets. The tolerance limit was set at
8 \masyr\ if the difference of the primary's Hipparcos and Tycho-2 proper motions
was larger than this value in at least one of coordinate components, and at 15 \masyr\ otherwise.
Additionally, the difference in proper motion between the primary and the candidate
companion was required to be within the 3-sigma formal error on the NOMAD proper motion.
These fairly strict limits removed from our analysis
some known or suspected nearby pairs, for example, the nearest stars Alpha and Proxima
Cen, which differ in proper motion by more than a hundred \masyr.
The resulting list of candidates was inspected by eye
to exclude numerous erroneous NOMAD entries. In this paper, we consider only CPM systems
with projected separations greater than 1000 AU.

Table~1 lists 41
CPM systems, including 2 resolved triple systems. Alternative names are given for all
companions, giving preference to Hipparcos numbers, various Luyten designations and
Gliese-Jahreiss identifications. The sources of J2000 ICRS positions are Hipparcos and
NOMAD. The $VI$ photometry comes mostly from \citep{bess, weis, wei93, wei96, koen, reid, ross}
and for several stars from our own observations. 

\section{Dynamical survival and origins}
\label{surv.sec}
Very wide stellar systems of low binding energy encounter other stars and molecular clouds as
they travel in the Galaxy, and these dynamical interactions are the main cause for stochastic
evolution of their orbits and, in most cases, eventual disruption. Analytical considerations
of dynamical evolution and survival of wide systems is limited to two asymptotic approximations,
those of very distant and weak (but frequent) interaction, and "catastrophic" encounters at
small impact parameters, which are rare but can be disruptive. We discuss in this paper the
second kind of interactions. According to \citet{wei}, catastrophic encounters are defined as 
those with impact parameters $b<b_{\rm BF}$, where $b_{\rm BF}$ is defined, in analogy with
Fokker-Planck diffusion coefficients, as the critical impact parameter at which the expected
variance of total energy is equal to a certain fraction of the total energy squared:
\eb
\sigma^2_{\Delta E}=\epsilon E^2.
\ee
The following proportionality relations are derived from \citep{wei} for the rates of catastrophic
encounters:
\begin{eqnarray}
\Gamma_{\rm cat} & \propto & n\,a\,\epsilon^{-1} \langle\frac{1}{V_{\rm rel}}\rangle 
\hspace{ 5mm} (b_{\rm BF}<<a) \label{enc1.eq}\\
\Gamma_{\rm cat} & \propto & n\,a^\frac{3}{2}\epsilon^{-\frac{1}{2}} \hspace{8mm} (b_{\rm BF}>>a)
\end{eqnarray}
where $n$ is the perturber number density, $a$ is the semimajor axis of the binary system,
$\langle\frac{1}{V_{\rm rel}}\rangle$ is the mean reciprocal relative velocity of encounters.

The first limiting case, $ b_{\rm BF}<<a$, corresponds to encounters with individual stars, while
the second, $ b_{\rm BF}>>a$, is a suitable approximation for encounters with dense cores of molecular
clouds. Note that in the latter case, the rate of high-energy interactions is independent of
the relative velocity.

The rate of disruptive interactions for both scenarios is proportional to the number density of
perturbers $n$. It becomes immediately clear that the rate of disruption of very wide binaries
is drastically different for the three major dynamical constituents of the Galaxy, the thin
disk, the thick disk and the halo. 

The halo stars populating the outer, spherical component of the Galaxy have by far the largest velocities
when they happen to travel in our neighborhood. The mean velocity with respect to the Local Standard
of Rest is directly related to the dispersions of velocity components $(\sigma_U,\sigma_V,\sigma_W)$.
The "pure" halo, according to \citet{chi}, is characterized by a prograde rotation of $V_\phi\simeq
30$ to 50 \kms, and a dispersion ellipsoid of $(\sigma_U,\sigma_V,\sigma_W)=(141\pm11,106\pm9,94\pm8)$
\kms. The vertical velocity component has immediate dynamical implications for wide binaries.
The number density of molecular clouds, as well as of field stars is non-uniform in the vertical
dimension, with a cusp at $z=0$. Wide binaries from the halo cross the densest part of the disk
where the chances of encounter are considerable very quickly, and spend most of their time
hovering far from the plane where the density of perturbers is much lower. On the contrary,
the thin disk stars spend most of the time within the densest parts of the Galaxy, oscillating
with small amplitudes around its midplane. These dynamical differences has dramatic implications
for the typical survival time of very wide binaries. We can quantify the differences in the
following way. 

According to the numerical simulations of galactic motion in \citet{makol}, the vertical oscillation
is harmonic to first-order approximation, with a period
\eb
P_\nu(v_{z0})\simeq P_{\nu,H}\,\left(1+\frac{|v_{z0}|}{10^4}(1.45+3.29\,|v_{z0}|)\right),
\ee
where $v_{z0}$ is midplane vertical velocity in \kms, and $P_{\nu,H}=77.7$ Myr is the
asymptotic harmonic period at $v_{z0} \rightarrow 0$. This equation holds within $\pm0.5\%$ for
$0\leqslant v_{z0}\leqslant 21$ \kms. Another useful equation relates the maximum excursion
from the Galactic plane with the midplane velocity:
\eb
z_{\rm max}=12.044\,|v_{z0}|.
\ee
Assuming typical midplane velocities to be equal to vertical dispersion estimates from \citep{tor}
for young stars, \citep{fam} for thin disk giants and \citep{chi} for the thick disk and halo,
we estimate characteristic midplane velocities, maximum vertical excursions, periods of
oscillation, and the fraction of lifetime spend in the dense part of the Galaxy for these
four dynamical components (Table~\ref{z.tab}). The latter parameter is defined as the
fraction of an oscillation period when the star is within 100 pc of the plane, $f(|z|<100)$.

The halo binaries cross the thin disk so quickly that their chances to encounter a perturber
(a field star or a molecular core) are relatively slim. Thus, generic binaries of very low
binding energy can probably survive for a long time in the halo. However, these objects are
rare in the solar neighborhood because of the intrinsic low number density, and none seem to be
present in our sample. A typical thick disk binary may also stay intact much longer than
a young disk binary, because it spends at least 6 times less time in the high-density 
midplane area. As far as encounters with stars are concerned, the difference in the time
of survival can be even greater, because the average reciprocal velocity of encounter enters
Eq.~\ref{enc1.eq}. Most of the interactions of thick disk binaries with thin disk perturbers
will be rapid, further reducing the rate of disruptive events. 

We can expect from this analysis that the distribution of very wide binaries and common proper
motion pairs in age should be bimodal. Young CPM pairs in the thin disk, despite the
higher rate of catastrophic interactions, can survive in significant numbers to this day.
This kind of binaries should be especially prominent if indeed most of the new stars are
born in loose comoving groups such as the Lupus association of pre-main-sequence stars
\citep{malup}.

\section{Color-absolute magnitude diagrams}
\label{hr.sec}
Fig. \ref{hr.fig} represents the joint $M_{Ks}$ versus $V-K_s$ color-absolute magnitude diagram for 
all resolved CPM companions
listed in Table~1 that have $V$ and $JHK$ magnitudes. We assumed in constructing this diagram
zero extinction for all stars, and we applied the Hipparcos parallaxes determined for
primary stars to their CPM companions, unless the latter have independent trigonometric 
parallax measurements. Known unresolved binary or multiple stars are marked with inscribed
crosses. A zero-age main sequence (ZAMS) and a 16 Myr isochrone at solar metallicity ($Z=0.001$)
from \citep{siess} are drawn with bold lines, and the empirical main sequence from
\citep{hen04} with thick dashed line. Some of the interesting stars discussed later in
this paper are labeled and named. Mutual position of primary and secondary CPM companions
are shown with dotted lines only for pairs with white dwarf companions.

The diagram shows that most of normal stars lie on or slightly above the main sequence.
This confirms that the fainter CPM companions are probably physical. We find that many of
the components lying close to the 16 Myr isochrone (upper bold line) are known visual, astrometric or
spectroscopic binaries, which accounts for their excess brightness. For example, the primary
component of the CPM pair {\bf HIP 66492} and {\bf NLTT 34706} is a resolved binary
with a period $P=330$ yr, semimajor axis $a=2.13\arcsec$ and eccentricity $e=0.611$ \citep{sey}. 
Formally, the joint magnitude can be as much as $0.75$ brighter (in a case of twin companions)
than the magnitude of the primary star. A number of components in Fig. \ref{hr.fig} lie
significantly outside the upper envelope of unresolved binaries defined by the empirical
main sequence minus 0.75 mag. Gross photometric errors (in particular, in $V$ for faint M dwarfs)
can not be completely precluded, but we suspect that most of these outlying stars should be
either very young or unresolved multiple stars.

The CPM pair of {\bf HIP 61451 and LTT 4788} is an emphatic example of overluminous stars
whose origin is an unresolved issue. They match the 16 Myr isochrone on the HR diagram
very well. The $K_s$-band excess for these companions is $0.7$ and $1.0$ mag, respectively.
We found no indication of binarity for either star in the literature. The primary component
can still be binary with an almost twin companion, but the secondary should be at least triple
to account for the near-infrared excess, if it is a normal (old) M2.5 dwarf. On the basis of
the kinematics of HIP 61451 and its excess luminosity, \citet{egg95} included it in his
list of the Pleiades supercluster, which is eponymous of the Local Young Stream \citep{makur}.
This may indicate an age between 1 and 125 Myr. Furthermore, HIP 61451 is a moderate and very
soft X-ray emitter (Table \ref{x.tab}), which may be expected of a post-T Tauri star. On
the other hand, its level of chromospheric activity at $\log R'_{\rm HK}=-4.601$ \citet{gra06}
is not impressive, corresponding to an activity age of $\simeq 1$ Gyr (see \S~\ref{age.sec}).
We propose that a careful investigation of the M-type  CPM companion LTT 4788 should
resolve the mystery of this system.

Fig. \ref{hr2.fig} shows a color-absolute magnitude diagram of some selected CPM components
discussed below in more detail, in $M_V$ versus $V-K_s$ axes. Each star is identified with
its Hipparcos number or other name. The two thin lines show the zero-age main sequence (ZAMS, lower)
and the 16 Myr isochrone (upper) from the models by \citet{siess}, both for $Z=0.001$ and zero
extinction. The thicker dash-dotted line is the empirical main sequence for field stars
from \citep{recr}. In this plot, the CPM components are connected with thin dashed lines
to show their relative position. Again, trigonometric parallaxes of the primary components
were assumed for faint companions with unknown distances.

\section{New CPM pairs}
\label{new.sec}
We report ten new possible CPM companions and 8 new CPM systems, including three at extremely 
large separations, netted out by our search
procedure (\S\ref{sel.sec}). Table~\ref{wds.tab} gives Washington Double Star catalog \citep[WDS; ][]{mas01}
identifications for the primaries of known systems and indicates new systems and wide
companions. The original discoverer references and other catalog identifications can be found in WDS.
In this paragraph, we discuss four probable new systems with peculiar characteristics,
which may be interesting to pursue
with additional photometric and spectroscopic observations.

{\bf HIP 109084} is a rather nondescript M0 dwarf at approximately 20 pc. This star has an uncertain parallax in
the Hipparcos catalog with a formal error of $7.9$ mas, most likely affected by unresolved binarity. 
According to \citet{giz}, the H$\alpha$ line
is in absorption (EW$=-0.55\AA$), hence the H$\alpha$ lower limit on age is
150 Myr (\S \ref{age.sec}). Its alleged CPM companion LP 759-25, as one of the nearest and latest M dwarfs, has
drawn more interest. \citet{pha} estimate a spectroscopic distance of 18 pc for this star.
At a projected separation of $65\,000$ AU, this may be one of the widest known CPM pairs, but
more accurate astrometric information is required to verify the physical connection between these
stars. 

The K3 dwarf {\bf HIP 4849} at 21 pc from the Sun is a binary resolved by Hipparcos and with 
speckle interferometry \citep{fab, bal}. Its inner companion is probably a K8 dwarf orbiting
the primary at $a=465$ mas with a period of 29 yr. We propose that this binary system has
a distant CPM companion, the DA5 white dwarf {\bf WD $0101+048$}. The projected separation between
the CPM components is $27\,000$ AU. The white dwarf companion is itself a binary star,
having a spectroscopically detected close DC white dwarf companion \citep{max}. The center-of-mass
radial velocity of the WD pair is $63.4\pm 0.2$ \kms, whereas \citet{nide} determine a
radial velocity of $22.17$ \kms\ for the primary K dwarf. The radial velocity for the WD
companion is likely to include the gravitational redshift, which may account for the apparent
difference with HIP 4849.  We estimate an age of $1.3$ Gyr for HIP 4849 from a chromospheric activity
index of $\log R'_{\rm HK}=-4.661$ given by \citet{gray}, whereas a cooling age for
WD $0101+048$ is $0.63\pm 0.07$ Gyr, not including main sequence lifetime \citep{ber}.

The pair of stars {\bf HIP 50564} and {\bf NLTT 23781} is remarkable not least because of its
extreme separation ($5230\arcsec$, or $111\,000$ AU on the sky). Other interesting properties
of this system are discussed in \S\ref{x.sec}.

\begin{figure}[htbp]
  \centering
  \includegraphics[angle=0,width=0.95\textwidth]{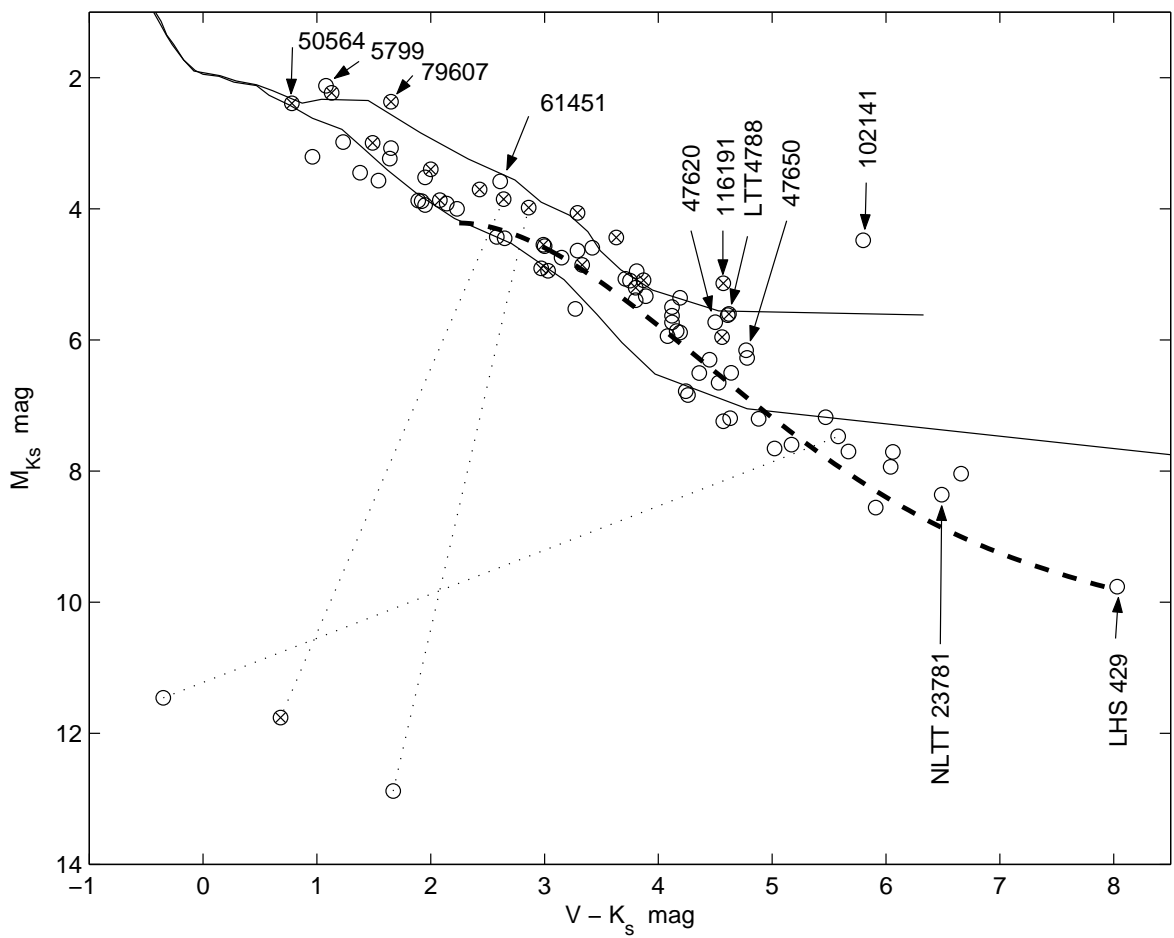}
  \caption{Joint color-absolute magnitude diagram of CPM pairs in $M_{Ks}$ versus $V-K_s$ axes.
The zero-age main sequence (ZAMS) and 16 Myr isochrone are drawn from \citep{siess}, both
for $Z=0.001$. The three white dwarfs of our sample are connected with their M dwarf companions
by dotted straight lines. The thicker dashed line indicates the empirical main sequence for
field dwarfs from \citep{hen04}. Known unresolved binary companions of all kinds are
marked with crosses inscribed in circles.}
  \label{hr.fig}
\end{figure}

The CPM pair of {\bf HIP 22498} (DP Cam) and {\bf G 247-35} is separated by "only"
$1\,000$ AU in the sky projection, and it is surprising it has not been identified as such before.
The primary component, a K7 dwarf, is listed as eclipsing binary in the catalog of eclipsing stars
\citep{malk}. Its Hipparcos parallax is very poor even for a "stochastic" solution,
indicating an unresolved type of binarity; however both this binary and its
distant M-type companion lie on the main sequence in Figs. \ref{hr.fig} and \ref{hr2.fig}. Very little is known
about G 247-35, apart from the photometric observations in \citep{wei88}.

\section{Activity and ages}

\label{age.sec}
\clearpage
\clearpage

\subsection{Chromospheric activity}
\label{chr.sec}
The so-called H$\alpha$ limit relation tells us that there is a certain age in the evolution
of M dwarfs of a given mass (or $V-I$ color) when the ubiquitous chromospheric activity, related
by emission in the H$\alpha$ line, disappears and the stars transform from
dMe to normal inactive dwarfs \citep{giz}. The empirical relation, fairly well defined on open clusters,
can be written as
\eb
\log\,{\rm Age}_{{\rm H}\alpha}=0.952(V-I_C+6.91).
\ee
This formula should be used with caution because recent studies of M dwarf activity based on
large samples of stars selected from the Sloan Digital Sky Survey indicate that the activity
lifetime versus spectral type relation is strongly nonlinear \citep{west08}, with a steep
ascent between M3 and M5. This abrupt change may be related to the transition from partially
convective to fully convective stellar interiors. Most of the latest M dwarfs in the Solar
neighborhood are active, but an age-activity correlation is still evident at spectral type M7
where the fraction of chromospherically active stars declines with the distance from the Galactic
plane \citep{west06}. This relation can be used to differentiate the oldest late-type M dwarfs,
although exact calibration is currently problematic because of the lack of independent age
estimates.

A widely used means of age estimation is provided by the empirical relation between the
level of chrospherical activity as measured from the $R'_{\rm HK}$ index of CaII lines.
The equation used in this paper,
\eb
\log {\rm Age}_{\rm HK} = (-2.02\pm0.13)\log R'_{\rm HK} - (0.31\pm0.63).
\label{hkage.eq}
\ee
derived by \citep{sode} for the Sun, Hyades and UMa Group. 
We utilize these relations to estimate (very roughly) the ages and age limits for several late type components
in Table~\ref{kin.tab}.

\subsection{X-ray activity}
\label{x.sec}
The binary and multiple systems under investigation in this paper are so wide that the observed
ROSAT sources can be unambiguously identified with individual components. Table~\ref{x.tab}
lists all the components identified by us in the ROSAT Bright Source and Faint Source catalogs
\citep{vog1,vog2}. The hardness ratios $HR1$ in this table are from the Rosat
catalogs, while the X-ray luminosities in units of $10^{29}$ \ergs\ are computed from the specified
count rates, hardness ratios and Hipparcos parallaxes. Most of the faint sources, with $L_X < 1$,
are very soft, with $HR1$ closer to $-1$. They are similar in X-ray activity to the quiescent Sun,
or slightly exceed it. The vast majority of weak nearby dwarfs are likewise soft, indicating
insipid coronal activity \citep{hun}. Normal M-type dwarfs have significantly smaller X-ray
luminosities than G- and K-type stars. Indeed, most of the X-weak systems include a K-type
primary, and a few F-type primaries, whereas numerous M-type wide companions are not detected by
ROSAT. A few notable M-type emitters should be mentioned.

The star {\bf HIP 14555} is a flare M0 dwarf with a Hipparcos parallax of $\Pi=52\pm 5$ mas.
This poor parallax determination, and a great deal of confusion associated with this multiple
system is related to a failed component solution in Hipparcos, based on the wrong assumption
that HIP 14555 and HIP 14559 (at separation $30\farcs3$, position angle $101\deg$) form a physical
pair at the same distance from the Sun. \citet{fama} resolved the Hipparcos data for this system
using more accurate initial assumptions, and obtained a parallax $\Pi=55.2\pm 2.5$ mas and a
proper motion $\mu=(-339,-121)\pm(2.5,2.2)$ \masyr\ for HIP 14555, which is quite close
to the original solution, but a $\Pi=8.8\pm 9.4$ mas and $\mu=(-18,-37)\pm(10,9)$ \masyr\ for HIP 14559.
Thus, these stars are certainly optical companions. {\bf LTT 1477} is probably a real,
albeit more remote, CPM companion. The outstanding X-ray brightness of HIP 14555 finds explanation
in the observations by \citet{giz} who find it to be a double-lined spectroscopic binary (SB2)
with a remarkable surface
velocity of rotation $v\sin\,i=30$ \kms. We are dealing with a typical extremely active M dwarf
in a multiple system: a short-period spectroscopic binary with a remote companion. The tertiary
companion may play a crucial role in the formation of the inner close pair via the Kozai cycle
and tidal synchronization of rotation \citep{kegg}, as discussed in \S\ref{kozai.sec}.

\begin{figure}[htbp]
  \centering
  \includegraphics[angle=0,width=0.95\textwidth]{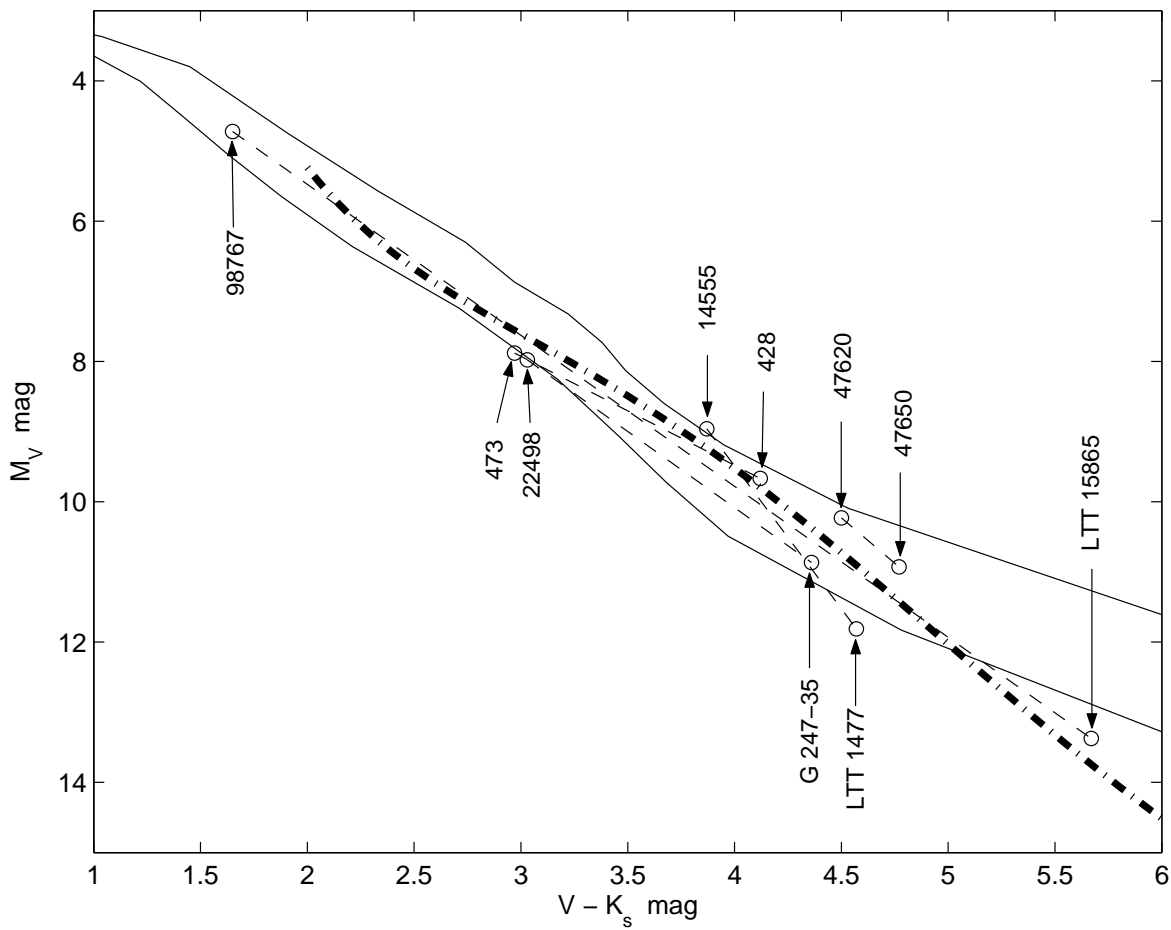}
  \caption{Color-absolute magnitude diagram of selected CPM companions in $M_{V}$ versus $V-K_s$ axes.
The zero-age main sequence (ZAMS) and 16 Myr isochrone are drawn from \citep{siess}, both
for $Z=0.001$. Components of CPM pairs are connected by
dashed straight lines. The thicker dot-dashed line indicates the empirical main sequence for
field dwarfs from \citep{recr}.}
  \label{hr2.fig}
\end{figure}

The star {\bf HIP 47650} is an M3 flare star and a member of the Hyades stream according to \citet{mont}.
\citet{nide} determined a "stable" radial velocity of $+6.6$ \kms for this star, precluding
a detectable spectroscopic companion. We should therefore consider the possibility that
this star is young. Both HIP 47650 and its brighter companion {\bf HIP 47620} lie significantly above
the empirical main sequence in Fig.~\ref{hr.fig}. These stars are brighter in $M_{Ks}$ than the
empirical main sequence from \citet{hen04} by 0.71 and 0.76 mag, respectively. Fig.~\ref{hr2.fig}
shows that both components are also brighter than their field counterparts in $M_V$ versus
$V-K_s$ axes as well, but by a smaller amount (0.49 mag in both cases). These photometric data suggest
a large $K-$band excess, probably due to a young age similar to the age of the Pleiades. The
substantial amount of X-ray radiation from HIP 47650 is accompanied by pronounced chromospheric
activity. According to \citet{wrig}, its average $S-$value of CaII chromospheric activity of
3.2 is outside and above the normal range where calibrated indeces $R'_{HK}$ can be estimated. 
What remains puzzling is that two stars of similar mass in a binary system can be so different
in chromospheric and coronal activity: HIP 47650 is a dMe star with EW$_{\rm H\alpha}=2.87\AA$ \citep{rauc},
whereas HIP 47650 has no emission in H$\alpha$ \citep{giz}. Since the difference in $V-I_C$ between the components 
is only 0.2 mag, employing formally the age criterion by \citep[][cf. \S \ref{chr.sec}]{giz} places the
system in very narrow brackets of age just above 1 Gyr. However, it seems unlikely that both
chromospheric and X-ray activity in the more massive companion HIP 47620 waned so abruptly;
the transition from dMe to normal M dwarfs is probably protracted and statistically uncertain.
This binary system indicates that the evolution of surface rotation, which is a crucial factor in solar-
and subsolar-mass dwarfs, may take different courses even for coeval, nearly identical stars. 

By far the brightest X-ray source in our collection is the BY Dra-type binary {\bf HIP 79607}
(TZ CrB, orbital period 1.14 d).
This example confirms that short-period spectroscopic binaries with evolved or solar-type primaries are
the most powerful emitters among normal stars, surpassing single pre-main-sequence stars
in X-ray luminosity by a factor of a few \citep{ma03}. The impressive flare activity on
this star was investigated in detail by \citet{oste}. Its distant companion HIP 79551 separated by at least
$13\,000$ AU is an M2.5 dwarf without any signs of chromospheric
activity; we surmise that it should be older than 3 Gyr (\S \ref{chr.sec}), setting a lower bound on
the age of the primary component. The primary, a F6$+$G0 pair of dwarfs \citep{fras}, has a visual 
companion at $5\farcs9$, orbital period 852.8 yr \citep{tok}. This inner companion ($\sigma1$ CrB)
may be responsible for the tight spectroscopic pair via the Kozai cycle, if the original orbits
were not coplanar (\S \ref{kozai.sec}). In this case, the substantial age of the system estimated
from the CPM companion is consistent with the time scale of dynamical evolution. The vertical
velocity component with respect to the Local Standard of Rest (LSR) is $+16$ \kms, assuming a
standard solar velocity of $W=+7$ \kms. This places the TZ CrB multiple system in the older
thin disk (Table~\ref{z.tab}, \S~\ref{surv.sec}), whose constituents spend roughly one third of
their lifetimes in the dense part of the Galactic disk. Thus, survival of the wide companion for
longer than 3 Gyr is plausible.

The star {\bf HIP 21482} appears to be another example of an extremely active BY Dra-type spectroscopic
binary in a hierarchical multiple system \citep{tok}. The inner spectroscopic pair has a 
orbital period of 1.788 d and is already circularized and rotationally synchronized \citep{mon97}. 
Its heliocentric motion (Table \ref{kin.tab})
is similar to the Hyades stellar kinematic group (SKG), except for the deviating, small $W$ 
velocity. The star was even suspected
to originate in the Hyades open cluster, which would fix its age at 600 Myr; in particular, \citet{egg93}
suggested that it can belong to the extended halo of evaporated stars around this cluster. The exceptional 
chromospheric activity of the inner pair at $\log R'_{\rm HK}=-4.057$ at the very tail of
the distribution observed for nearby field stars \citep{gray}, may also indicate a young age.
However,
the remote companion {\bf WD 0433+270} is a cool DC white dwarf, and therefore, the system can hardly be young.
\citet{ber} estimated a $T_{\rm eff}=5620\pm110$ K and cooling age of $4.07\pm0.69$ Gyr, an
order of magnitude older than the Hyades. In this case again, the WD companion had plenty of time
to tighten and to circularize the inner pair via the Kozai cycle and dissipative tidal friction. 

The star {\bf HIP 115147} (V368 Cep) is one of the nearest post-T Tauri stars. It is mistakenly identified
as a RS CVn-type binary in the Simbad database, although, contrary to the previously discussed
objects of this type, it is not a spectroscopic binary. Both its secondary companion at $11\arcsec$
and the newly discovered tertiary CPM companion LSPM J2322+7847 \citep{makza} lie significantly
above the main sequence in optical and infrared colors. The probable age of this system is only
20--50 Myr, and the high rate of rotation of the primary (period 2.74 d, \citet{kah}) is obviously
due to its youth. The origin of this post-TT triple system is an open issue, a high-velocity ejection
from the Ophiuchus SFR being one of the possibilities considered.

The pair of outstanding T Tauri stars {\bf HIP 102409} (AU Mic) and {\bf HIP 102141} (AT Mic) epitomize the
class of very young, active X-ray emitters. They may be as young as 10 Myr, and both display the
whole complement of stellar activity indicators. AU Mic has a nearly edge-on debris disk, and
its remarkable X-ray luminosity is probably nurtured by the high rate of rotation with a period
of surface spots of 4.847 d \citep{hebb}. Its distant companion, AT Mic, is a flare M4.5 dwarf
and an extreme UV source. Both stars lie significantly above the 16 Myr isochrone in Fig.
\ref{hr.fig}. AT Mic has a somewhat poorly investigated companion LTT 8182 at 3.8 arcsec,
position angle $218\degr$ which is missing in the 2MASS survey and omitted in Table~\ref{big.tab}.
Its H$\alpha$ emission is also remarkably high (EW$=9.3\AA$, \citet{scho}).
AT Mic and AT Mic are separated by more than 0.2 pc in the sky plane, one of the largest separations
found in this paper, and it is unlikely the two stars can be gravitationally bound. They will
inevitably part their ways in the future, as well as other members of the dispersed BETAPIC
stream \citep{ma07}.

\clearpage
\clearpage

Both components in the CPM pair {\bf HIP 25278 and 25220} are prominent X-ray sources \citep{hun}.
The primary GJ 202, an F8V star, is however more than 10 times weaker than its K4 companion GJ 201.
Both stars have been assigned to the Hyades SKG by \citet{mont}. HIP 25278 appears to be a single
star of slightly subsolar metallicity with an estimated age of 5.6 Gyr \citep{nord}. \citet{taka}
determine a slightly higher [Fe/H]$=0.05$ and find a surprisingly high content of Lithium (EW$=0.094\AA$).
Another unexplained characteristic of this star is its position below the main sequence in
Fig. \ref{hr.fig}. The moderate X-ray activity is accompanied by a noticeable CaII chromospheric
signature at $\log R'_{\rm HK}=-4.38$ and rotation $P/\sin i = 4.1$ d  \citep{rein}. Using the
above value for $\log R'_{\rm HK}$ and Eq. \ref{hkage.eq}, we obtain an age of 0.3 Gyr (Table \ref{x.tab}),
significantly less than Nordstr{\"o}m et al.'s estimate, and roughly consistent with the age of
the Hyades open cluster. The CPM companion GJ 201, an active K4V star, has a $\log R'_{\rm HK}=-4.452$
\citep{gray}, and hence, an age of 0.48 Gyr. It appears to be spectroscopically single. Its Lithium
abundance is low, however \citep{fav}. Furthermore, the H$\alpha$ line is in absorption according
to \citet{herb}, placing this star in the realm of inactive, regular dwarfs. The high level of
X-ray activity in this stars remains a mystery, because it can not be explained just by the relative
youth. Indeed, the distribution of X-ray luminosity between the companions appears to be inverted
to that observed in the Hyades cluster \citep{ster}, in that the F8 primary companion is below the
lower envelope of $L_X$ for its Hyades counterparts, while the secondary component, GJ 201, is roughly
a factor of 10 more luminous than the average K dwarf in the Hyades, and is comparable in X-ray
emission to the brightest non-binary F8-G0 Hyades members.

The stars {\bf HIP 116215} and {\bf 116191}, of spectral types K5 and M3.5, respectively, have
space velocities similar to the Local stream of young stars \citep{mont}. They may be as young as
the Pleiades. The primary component (GJ 898) is single and its X-ray luminosity is similar to the average
value for the Hyades late K-dwarfs. The secondary (GJ 897 AB) is a resolved visual binary \citep{maso}
with an orbital period of 28.2 yr and a semimajor axis of $0.59\arcsec$, which probably explains
why this M dwarf lies significantly above the main sequence, while the primary is quite close to it.
A $\log R'_{\rm HK}=-4.486$ from \citet{gray} for HIP 116215 implies an age of 560 Myr, again similar to
the age of the Hyades. The H$\alpha$ line is in absorption for the primary, but prominently in
emission for the secondary \citep{giz}. This fact can be used to estimate the boundaries of
H$\alpha$-age (\S \ref{chr.sec}), which yields $\log$(Age)$\in[8.1,9.0]$, in good agreement with the
$\log R'_{\rm HK}$-age estimate. The remaining difficulty in the interpretation of this system is the
unusual strength of X-ray emission from HIP 116191, by far surpassing the levels observed for this
age and spectral type in the Hyades. One may suspect that one of the visual companions in this
binary is an undetected short-period spectroscopic binary.

The star {\bf HIP 46843} is likely another representative of young X-ray emitters. A $\log R'_{\rm HK}=-4.234$
from \citet{gra06} yield an age of 175 Myr, in fine agreement with the rotational age estimate
164 Myr from \citet{barn}. Its M5.5 companion {\bf GJ 9301 B} is undetected in X-rays. The young age
of this system is confirmed by the $L_X$ in Table~\ref{x.tab} for HIP 46843, which is only slightly
smaller than the typical luminosity of Pleiades members \citep[$\sim 3\times10^{29}$][]{stau} of
this spectral type. GJ 9301 B is therefore one of the youngest late M dwarfs in the solar
neighborhood. Note that Simbad mistakenly provides an uncertain estimate of $M_V$ from \citep{reihg}
as a $V$ magnitude.

The star {\bf HIP 50564} of spectral class F6IV is remarkably active in X-ray but is unremarkable
chromospherically \citep[$\log R'_{\rm HK}=-4.749$;][]{gray} and depleted in lithium. The low degree
of activity points at an age of 1.9 Gyr. On the other hand, this star is a $\delta$ Scuti-type
variable and a fast rotator, $v\sin\,i=17$ \kms. It has a solar iron abundance, [Fe/H]$=0.09$
from \citep{nord} and a space motion typical of the local young stream, $(U,V,W)=(-14,-26,-12)$ \kms.
The key to the mystery of its X-ray activity may be in a short-period, low-mass companion;
indeed, \citet{cuti} mention that the star is "reported as SB1" (single-lined
spectroscopic binary) without providing further detail.
Its M5-type CPM companion {\bf NLTT 23781} separated by at least 0.5 pc is one of the discoveries in this paper.
It was cataloged in \citep{lepi,sal}, but otherwise, this interesting object completely escaped the
attention of observers. Its location in the HR diagram (\ref{hr.fig}) above the main sequence
indicates a young age or binarity. Thus, this extreme system represents a mystery in itself.
If it is indeed 1.9 Gyr old, how could it survive at this separation having spent all the time
in the thin disk, and why the remote companion is overluminous?

The star {\bf HIP 59000} has a known CPM companion {\bf NLTT 29580} separated by $4\,200$ AU in the sky
projection. \citet{gra06} report a substantial chromospheric activity of the primary,
$\log R'_{\rm HK}=-4.341$, which translates into a chromospheric age of 0.29 Gyr. HIP 59000
is orbited by a low-mass companion, probably a brown dwarf, for which we derive a first orbital solution
in \S\ref{bin.sec}. This inner companion is not close enough to the primary ($P\simeq 5.1$ yr) to
account for the significant X-ray luminosity of the system. We think that either the primary is
a yet-undetected short-period spectroscopic binary (in which case the astrometric companion
may have a stellar mass), or the system is indeed fairly young. The remote CPM companion
NLTT 29580, a M5.0 star, is confirmed the photometric parallax from \citet{reid03} being in excellent
agreement with the updated parallax of HIP 59000 (45 mas).

There is little doubt that the origin of the X-ray activity in {\bf HIP 82817} is in the innermost
component of this intriguing system of at least five stars, which drives the fast rotation of the
secondary. Indeed, the A component is orbited by a B component at $P_{\rm AB}=626$ d, which is in fact some 50\%
more massive than the primary because it is a spectroscopic binary with a period
$P_{\rm B}=2.96553$ d and a mass ratio of 0.9 \citep{maz}, made of nearly identical M dwarfs. 
Both eccentricities are low, and the orbits are likely to be coplanar. The widest CPM companion
{\bf LHS 429} is a M7 dwarf lying on the empirical main sequence for late field dwarfs (Fig. \ref{hr.fig}).
\citet{maz} suggest an age of $\sim5$ Gyr for the system.

The F5V star {\bf HIP 5799} and its G9 CPM companion {\bf GJ 9045 B} are moderately metal-deficient ([Fe/H]$=-0.3$),
kinematically belong to the thin disk population and have an estimated age of 2.5 Gyr \citep{soub}.
This age estimation is supported by the moderate HK activity obtained by \citet{gray} for the primary.
The combination of a significant X-ray emission from the primary and the lack of such from the
secondary, a modest rotational velocity of HIP 5799 \citep[$v\sin\,i=4.4$ \kms;][]{tok90} and
the above age are puzzling. The peculiar location of HIP 5799 in the HR diagram (Fig. \ref{hr.fig})
much above the main sequence and closer to the 16 Myr isochrone may give a clue. This star may be
a yet undetected short-period spectroscopic binary seen almost face-on.

The CPM pair {\bf HIP 86036} (= 26 Dra) and {\bf HIP 86087} (= GJ 685) represents a rare case
when both components are detected by ROSAT. Their X-ray luminosities differ by more than a factor of 10
which may be the natural consequence of the difference in the sizes of their coronae, the primary
being a G0V star and the distant companion a M1V dwarf. The primary is in fact a triple system where
A component has a 76 yr orbiting B companion and a wide low-mass C companion at $12.2\arcsec$
\citep{tok}. Definitely, these resolved companions (not present in our sample) are not responsible
for the enhanced X-ray emission from the inner system and we have to look for signs of a young
age. Surprisingly, we find conflicting data. The primary star HIP 86036 is moderately metal-poor ([Fe/H]$=-0.18$)
and has an age of 8.4 Gyr according to \citet{soub}. \citet{nord} give an even older age of
11.5 Gyr for this star. However, the rotational age of the distant companion GJ 685 is only
$435\pm50$ Myr at $P_{\rm rot}=18.6$ d \citep{barn}. The H$\alpha$ line is in absorption 
\citep[EW$=-0.4\AA$;][]{stau}, which only means that this M1V star is probably older than
200 Myr. Another confusing detail comes from the CaII HK line flux which is low for this type
of star and the period of rotation \citep{rutt}. It is possible that the fast rotation of
GJ 685 is driven by extraneous agents, and the rotation age estimate is confused. 
To summarize, the origin of X-ray activity and
the age of this system remains unknown.

\section{Multiplicity}
\label{bin.sec}

At least 17 out of our 41 CPM systems contain inner binary or triple components. We have reasons to
believe that some of the CPM components are still undiscovered binaries, especially those
stars that are too luminous for their spectral type and age, and have enhanced rates of
rotation and chromospheric activity. The rate of triple and higher-order multiple systems
among non-single stars in the Hyades is only 0.14 \citep{pati}, significantly smaller than
we find for CPM pairs (0.41). To some extent, the high-order multiplicity of very wide
pairs can be explained by the higher mass of binary stars and therefore, better chances of
survival in the course of dynamical interaction with other constituents of the Galaxy. This
argument may be particularly relevant for older CPM systems of extreme separations. On the
other hand, there may be a more subtle reason for the abundance of hierarchical systems.
The primary fragmentation of a prestellar molecular cloud and the secondary fragmentation during H$_2$
dissociation are likely to take place at two distinct hierarchical spatial scales \citep{whit}.
Of the three main models of low-mass star formation considered in that paper, the 2D
fragmentation triggered by supersonically colliding gas streams appears to be the most plausible
scenario for wide companions in multiple systems. It predicts a wide range of initial
orbital eccentricities and relative inclinations in such systems.

Perhaps the system of CPM companions {\bf HIP 473} and {\bf 428} is the most important for empirical
study of the Kozai-type evolution of multiple systems. The latter star, an M2e dwarf, is known
as the F components of the system ADS 48, where the primary star has a visual twin companion B (spectral
type M0) separated by $6\arcsec$. The most interesting aspect of this system is
that the mutual inclination of the B and F companions is $\simeq 80\degr$ according to the
family of probable orbits computed by \citet{kiy}. The eccentricity of the inner pair AB is
probably between 0.2 and 0.6. Therefore, ADS 48 may be a paragon of the Kozai evolution in progress, where
the inner pair has not yet shrunk but remains in an elliptical orbit. \citet{kiy}
note a probable inner tertiary companion, which may account for the total dynamical mass higher 
by $\sim 0.3$ \msun\ than what is expected from the spectral type. Furthermore, they note a
slight variation in position of the A component with a period of 15 yr, possibly indicating another
$\simeq 0.05$ \msun\ companion. The A component lies on, or slightly below, the main sequence
in Fig. \ref{hr2.fig}, thus, the hypothetical companions contribute little in the total
luminosity. \citet{ano} pointed out that the probability of a hyperbolic orbit for the F
component appears to be greater than of an elliptical orbit. Such systems may be unstable in the long run.
In the latter paper, it is proposed that ADS 48 is a member of the Hyades flow, of which we have quite a few
representatives in our selection (Table \ref{kin.tab}). Stars in a kinematically coherent stream
are more likely to be found in accidental slow passages near each other. The star HIP 428 is
an emission-line M2 dwarf \citep{rauc}, indicating an upper limit of $\simeq 1$ Gyr on age. This estimate
is consistent with the upper envelope of the Hyades flow \citep{egg98}. We believe that 
Kiyaeva et al.'s suggestion that the F companion is physically bound to the AB pair with an
orbital period of $\sim 10^5$ yr is more plausible in the light of recent astrometric data.

The nearby star {\bf HIP 14555} (GJ 1054A) along with its optical companion HIP 14559 epitomize the
difficulties that arise in the reduction of Hipparcos data for visual multiple systems (\S\ref{x.sec}).
The improved solution for HIP 14555 from \citep{fama} is $\Pi=55.5\pm 2.5$ mas, $(\mu_\alpha\cos\delta,\mu_\delta)=(-339,-121)$ \masyr, which is close to the original results. The remaining inconsistency
is that with the estimation by \citep{henr02} who inferred a distance of 12.9 pc based on their
spectral type determination and the $V$ magnitude specified in Hipparcos. This biased estimate comes
from the photometric data which seem to be too bright. Fig.~\ref{hr2.fig} depicts the HR diagram
for both HIP 14555 and the alleged CPM companion LTT 1477, with photometric data from \citep{wei93} and
the same parallax of $55.5$ mas assumed for both stars. The primary component lies significantly
above the empirical main sequence and appears to match the 16 Myr isochrone from \citep{siess}.
Despite the prominent H$\alpha$ emission and X-ray activity, this star is not considered to be
young. The apparent brightness excess is the consequence of unresolved binarity of HIP 14555. Indeed,
according to \citep{giz}, the star is double-lined spectroscopic binary (SB2). 

The star {\bf HIP 34052} (GJ 264) is the tertiary component of a well-known wide triple system, which also
include the pair of solar-type stars GJ 9223 (A) and GJ 9223 (B), separated by $21\arcsec$ on the sky.
By virtue of the high proper motion and brightness, the system has been included
in the lists of nearby star for a long time, attracting considerable attention due to the
possibility of testing the evolution of stellar gravity, temperature and chemical composition
in great detail. The spectroscopic investigation of components A and B by \citet{chmi} found a common iron
abundance of [Fe/H]=$-0.27\pm0.06$ and effective temperatures $5870\pm40$ K and $5290\pm70$ K,
respectively. Somewhat different lithium abundances were determined for the two components, but
both at the solar level or below it. These estimates, together with the Galactic orbit (eccentricity
0.31) and a negligible chromospheric activity from the \ion{ca}{2} lines, indicate an old system,
probably representing the old disk. A theoretical ZAMS used by \citet{chmi}, adjusted to the
location of the B component on a $\log T_{\rm eff}$--$M_{\rm bol}$ diagram yielded a parallax of
$68$ mas. The trigonometric parallax of the system is close to $60$ mas (Table~\ref{big.tab}). It may
be suspected that the B component is too bright for the estimated $T_{\rm eff}$ and metallicity.
However, all three companions lie close to the main sequence with their photometric parameters
in Table~1. 

Relatively little is known about the tertiary component, HIP 34052. A robust astrometric solution
was produced for this star in Hipparcos, without any indications of binarity or variability.
However, the Hipparcos proper motion $(\mu_\alpha\cos\delta,\mu_\delta)=(-75.4,401.3)$ \masyr~
differs significantly from the Tycho-2 proper motion $(-93.0,395.3)$ \masyr\ \citep{hoeg}. Since the
latter is systematically more accurate in the presence of orbital motion, \citet{makap} included
it in the list of astrometric binaries with variable proper motion. We further elaborate on this star
by applying a multi-parameter orbital optimization algorithm designed for the Hipparcos Intermediate Astrometry Data (HIAD) \citep[see, e.g.,][]{mavan}. This algorithm, based on the Powell method of nonlinear
iterative optimization, looks for the global minimum of the $\chi^2$ statistics on abscissae residuals
specified in the HIAD, corresponding to a certain combination of 12 fitting parameters, including
seven orbital elements and five astrometric corrections. 
The estimated orbital parameters are: period $P=1501$ d, inclination $i=180\degr$, apparent semimajor
axis $a_0=30.6$ mas. The formal F-test on reduced $\chi^2$ ($0.933$ after orbital adjustment) equals
$1.0$. The orbit is incomplete, because the period is longer than the time span of Hipparcos observations.
Therefore, the orbital elements are fairly uncertain, and follow-up observations are needed to
estimate the mass of the system. Assuming that the total mass of the system is $1.0$ \msun, the
companion mass is only $0.2$ \msun, and the angular separation is about 150 mas ($a=2.6$ AU).
The companion may be possible to resolve with the HST or ground-based coronographic
facilities.

Astrometric binarity of the {\bf HIP 59000} (GJ 9387) K7 dwarf is advertised by its varying 
proper motion \citep{makap}.
It is not a known spectroscopic binary; therefore, we attempted an unconstrained 12-parameter
orbital solution for this star using the same algorithm described in the previous paragraph.
A visual inspection of the HIAD data reveals that the orbital period is several years, and
we are dealing with another incomplete orbit. As a consequence, the fitted parameters should be
considered preliminary. We obtain a period $P=1854$ d, apparent semimajor axis $a_0=12$ mas, $T_0=$JD
2448368, $\omega=61\degr$, $\Omega=53\degr$, inclination $i=74\degr$ and eccentricity $e=0.6$.
The updated parallax is $\Pi=45.5\pm0.7$, which is close to the original Hipparcos parallax.
The standard error of $a_0$ is about 2 mas, but the eccentricity is quite uncertain. Assuming
a mass of $0.5$\msun~ for the visible primary, its apparent orbit on the sky leads to a total
$a=2.34$ AU and a secondary mass of $0.063$ \msun. The expected radial velocity semi-amplitude
is $1.9$ \kms. Thus, this newly discovered binary system contains a brown dwarf which may be
only 290 Myr old (\S \ref{chr.sec}).

The star {\bf HIP 75718} (GJ 586 A) is the primary in a system of at least four components 
\citep{tok}. The system is enshrouded
in puzzles. The inner pair is both spectroscopic and astrometric \citep{duqu, janc}
yielding a fairly detailed orbit. It consists of a K2V dwarf (mass 0.74 \msun) and a later K dwarf
(mass 0.49 \msun) in a $889.6$-day orbit. The orbit has an outstanding eccentricity of
$0.9752\pm 0.0003$, so that the separation between the companions at periastron is only
about 10 solar radii. The tertiary companion {\bf HIP 75722} $=$ GJ 586 B is separated by $52\arcsec$
in the sky projection. It is another K2V dwarf of the same mass as the primary of the inner pair
(0.74 \msun). \citet{hun} assign the considerable X-ray flux detected by ROSAT to both A and B
components, but in our opinion, it is the B component, surprisingly enough, that is responsible
for the X-ray source (Table \ref{x.tab}). The two companions are disparate in their CaII line
activity too, the A component being at $\log R'_{\rm HK}=-4.97$ indicating an old star, and the B
at $\log R'_{\rm HK}=-4.37$ \citep{wrig}. Formally, we would estimate the chromospheric age
(\S \ref{kozai.sec}) at 0.33 Gyr. Furthermore, the A and B components have different rates of
rotation, $P_{\rm rot}=39.0$ and $9.0$ d, respectively. What could be the reason for the high
activity and fast rotation of GJ 586 B? \citet{tok91} reported outlying radial velocity measures
(spikes) for this star in otherwise constant series of observations and suggested that the B
component can also be a high-eccentricity spectroscopic binary. If this is the case and the
orbital period is of order a few days, the discrepant activity levels and age estimates are explained.
However, \citet{nide} report a constant radial velocity from their extensive measurements.
To confuse the matter more, \citet{nord} specify a fairly low probability ($0.285$) of constant
radial velocity from their 18 observations spanning 6014 d. We consider the system of
HIP 75718 and 75722 to be one of the best targets to investigate the Kozai cycle in action.
It is not clear if the tidal friction in the primary inner pair is sufficient to tighten it up,
but the secondary may prove to have dynamically evolved under the influence of the primary.
Finally, it is not clear if the mysterious fourth component GJ 586 C (G 151-61) is physically associated
with this triple or quadruple system. Its trigonometric parallax \citep{dahn82} is tantalizingly
close ($\Pi=47\pm5$ mas), but the proper motion is $\sim 15\%$ smaller. NOMAD supplies us with
the following data for this star: position J2000 $15\;27\;45.08$, $-9\;01\;32.5$, proper motion
$\mu=(30,-312)\pm(2,3)$ \masyr. The available magnitudes are $V=15.41$, $J=10.55$, $H=9.92$ and
$K_s=9.63$. The smaller proper motion of this late M dwarf accounts for its absence in our
NOMAD-based sample. Since the system appears to be genuinely old, it is doubtful that GJ 586 C
can form a kinematic group with the brighter counterparts.

\subsection{Candidate stars with planets}
\label{kozai.sec}
It is commonly accepted that planets can be present in binary stellar systems. The latest
investigations in this area indicate that 23\% of candidate exosolar planetary systems also have stellar
companions \citep{ragh}. Planets can form in stable circumbinary disks if the latter are large enough,
so that the stellar binary and the distant planet form a dynamically stable hierarchical
system. In very wide CPM systems, we encounter a different hierarchical composition, when
the remote tertiary companion has a stellar mass. Such planetary systems may be subject to
the long-term oscillatory perturbations of inclination and eccentricity 
over a long time (several Gyr) because of the secular loss of orbital energy known as the
Kozai cycle. The eccentricity variation is especially important for the dynamical
evolution of the inner planetary system. The Kozai-type variation is significant only
if the tertiary companion has a different initial inclination from the inner orbit \citep{malm}.
If, for example, the initial inclination of the tertiary is $76\deg$, the planet will
periodically describe an orbit of $e=0.95$. This high eccentricity entails very close
periastron passages of the primary. Giant gaseous planets will be subject to the
tidal friction at periastron passages quite similar to the mechanism suggested for
stellar binaries. The gradual loss of angular momentum may lock the planet on a
high-eccentricity orbit, resulting in secular shrinkage of the orbit. The orbits of very short-period
`hot jupiters' ($P<10$ d) should be circularized similarly to tight spectroscopic binaries.
It is also important to note that the dynamical evolution due to the Kozai mechanism may
be quite different for single planets and stable planetary systems even if the initial
inclination of the tertiary stellar companion is high. \citet{inna} point out that a system
of four major Solar System planets would remain stable and roughly coplanar in the presence of
a distant companion on timescales much longer than the timescale of the Kozai cycle,
owing to the mutual dynamical interaction between the planets.

Our sample of CPM systems includes two candidate exoplanet hosts. The star {\bf HIP 43587} (GJ 324,
55 Cnc), which has a co-moving companion {\bf LTT 12311} at a projected separation of 1050 AU,
is a solar-type dwarf suspected of bearing a system of at least four planets \citep{maca}.
One of them (55 Cnc d) is a super-Jupiter with a mass $M\sin i=3.9M_{\rm J}$, a period of
about 5550 days and a semimajor axis of nearly 6 AU. The other three suggested planets have
masses between 0.037 and 0.83 $M_{\rm J}$ and periods ranging 2.8 to 44 days. The spectroscopically
determined eccentricities are all small ($<0.1$). There are a few conflicting clues about
the age of the stellar components. The star 55 Cnc lies above the
empirical main sequence by 0.55 mag according to \citep{butl}. Both this star and its
companion GJ 324 B lie slightly above the main sequence in the $M_{Ks}$ versus $V-K_s$
diagram in Fig.~\ref{hr.fig}. The primary has a moderately enhanced metallicity [Fe/H]$=0.315$,
common among exosolar planet hosts. However, the chromospheric activity of 55 Cnc is quite
low, at $\log R'_{\rm HK}=-5.04$ \citep{wrig}, which is in fact close to the mean chromospheric
flux parameters for the most inactive field solar-type dwarfs \citep{gray}. Wright et al. 
estimate a $\log$(age) $=9.81$, and indeed, a similar age of 9.87 (7 Gyr) is obtained from the
HK index. \citet{mont} list 55 Cnc as a member of the populous Hyades stream (or kinematic group), based
on its heliocentric velocity vector (cf. Table~\ref{kin.tab}). In the light of recent
investigations, the Hyades stream, originally believed to originate from the evaporating
Hyades supercluster \citep{egg93} of approximately 700 Myr of age, incorporates stars of
a wide range of age and chemical composition, indicating a curious phenomenon of dynamical
alignment \citep{fam}.

The star {\bf HIP 98767} (GJ 777 A) is similar to 55 Cnc in metallicity ([Fe/H]$=0.213$) and brightness excess
($\Delta M_V=0.66$) according to \citet{butl}. Both this star and its distant CPM companion
{\bf LTT 15865} separated by more than 2800 AU, lie slightly above the main sequence in Fig.
\ref{hr.fig}, but perfectly on the empirical main sequence in Fig. \ref{hr2.fig}. The primary star 
lies above the main sequence by 0.66 in absolute $V$ magnitude according to
\citep{butl} and is moderately metal-rich, [Fe/H]$=0.213$. We do not know how to interpret
these discordant photometric data, except to assume that there is some anomaly in the
$B$ and $K$ bands. The primary is suspected to 
bear not just a single planet but
a system of at least two planets \citep{vo5},
a short-period HD 190360 c of mass $M\sin i \approx 0.06 M_J$ and $P=17$ d, and again a Jupiter-like
HD 190360 b of mass $M\sin i \approx 1.55 M_J$ and orbital period 2900 days. The eccentricities are
$\approx 0$ and 0.36, respectively. It is possible that the inner planet has already been circularized
by tidal friction, whereas the outer massive planet is still undergoing its Kozai cycles.

\section{Moving groups and streams}
\label{skg.sec}
The solar neighborhood is permeated with stellar kinematic groups (SKG), which are evident as number
density clumps in the 3D velocity space \citep{cher}. Since these streams are only loosely coherent
kinematically, and are not supposed to be gravitationally bound, their existence poses a
certain problem of dynamics. The Hyades stream figures prominently in our sample (Table \ref{kin.tab}).
The Sun is located inside the Hyades stream today (but does not belong to it), so that any
selection of the nearest stars will give preference to this SKG, as opposed to, for example,
the Ursa Major SKG. Still, the large number of CPM systems in the Hyades stream is surprising
for the following reason. Recent investigations indicate that the Hyades SKG is composed of
stars and clusters of disparate ages and origins \citep{fam}, contrary to the previous hypothesis
that it is the result of dynamical evaporation of a massive open cluster. But if
this stream is purely dynamical phenomenon, a kind of focusing taking hold of random unrelated
objects, why do we find so many CPM pairs which are apparently generic? A dynamical agent
sufficiently powerful to rearrange the local 6D phase space of the Galaxy would probably
accelerate the disruption of wide binaries rather than preserve them. Furthermore, the
possible Hyades stream members present in our sample do not look like random field stars.
Many of them have enhanced levels of chromospheric and X-ray activity indicative of moderately
young age ($\sim 1$ Gyr, roughly consistent with the age of the Hyades open cluster).
On the other hand, the presence of weakly bound CPM pairs in the Hyades SKG is not consistent
with the dynamical evaporation paradigm, because the latter assumes a dynamical relaxation
and ejection event. A pair of M dwarfs like HIP 47620 and 47650 (discussed in
\S\ref{x.sec}) is unlikely to be thrown out of the Hyades cluster and remain intact.

The star {\bf HIP 4872} and its distant companion {\bf V388 Cas} (GJ 51) are related to
the Hyades stream by \citet{mont}. The latter companion is a well known M5 flare dwarf
of considerable X-ray luminosity (Table~\ref{x.tab}) and EUV activity \citep{chri}.
A better age estimate can be obtained for the former companion which is a flare M1.5 dwarf. 
\citet{rauc} list this star as dMe with an H$\alpha$ equivalent width of $2.0\AA$.
This yields an upper age limit of 280 Myr. The young age and the activity levels are
consistent with this system being in the young core of the Hyades flow.

The pair of CPM companions {\bf HIP 15330} and {\bf 15371} ($\zeta^1$ and $\zeta^2$ Ret)
is remarkable because both stars lie significantly below the ZAMS for $Z=0.01$.
They could be suspected to be metal-poor, but the iron abundance is only moderately
low at [Fe/H]$=-0.22\pm0.05$ according to \citet{delp}. The pair was originally assigned to the
$\zeta$ Her SKG, but since the latter star does not appear to belong to the moving group,
it was renamed to $\zeta$ Ret SKG. Lately, \citet{soub} determined somewhat smaller
iron abundances ($-0.34$ and $-0.30$) for the two stars and assigned them to the Hercules
SKG. The vertical velocity is $W=16$ \kms, the maximum excursion from the plane $z_{\rm max}
=0.31$ kpc, and the eccentricity of the Galactic orbit $e=0.26$. \citet{all} propose to
define the thick disk as either $e \geq 0.3$ or $|z_{\rm max}| > 400$ pc. Thus,
the CPM pair in question does not qualify for the thick disk by any of these kinematic
criteria. The origin of this system and its peculiar blueness remains an unresolved issue.

\section{Discussion}
\label{dis.sec}
One of the most interesting results of this paper is that we find little, if any, presence
of thick disk or halo population in the local sample of very wide binaries. The only
CPM system that may belong to the thick disk is the WD+dM4.5 pair HIP 65877 (DA3.5
white dwarf WD $1327-083$) and LHS 353, cf. \citet{sil}. This
shows that even the widest pairs at separations greater than $1000$ AU can survive for
$\simeq$ 1 Gyr staying constantly in the thin disk of the Galaxy, despite numerous
encounter and dynamical interaction events. This observation does not refute the
dynamical analysis presented in \S~\ref{surv.sec}, because thick disk and halo stars are
very rare in the Solar vicinity, and our sample (based on bright stars in the Hipparcos
catalog) is probably too small and incomplete to accommodate
sufficient statistics. But if we boldly extrapolate this result to a wider part of the
Galaxy, we conclude that normal thin disk, moderately young or very young, stars dominate
wide CPM binary and multiple systems. Statistically, this is quite consistent with the
estimation by \citet{bart} on a larger sample of 804 Hipparcos visual systems,
who found that $92\%$ of systems belong to the thin disk (and are mostly young to middle-age),
$7.6\%$ to the thick disk, and much less than $1\%$ to the halo. Further inroads in this
study can be made by collecting a larger volume-limited sample of very wide binaries and
a comparison with a representative set of nearby field stars. We consider this paper as an initial
step in this direction.

Despite the considerable progress in recent years, chronology of solar-type stars is
still in a rather pitiful state, and we find more evidence to this in the widely
discrepant age estimates for a few CPM systems obtained with different methods.
Although a significant fraction of CPM companions display enhanced chromospheric,
X-ray and EUV activity, only few are patently in the pre-main-sequence stage of evolution
(e.g., AT and AU Mic) where these signs of activity and the high rate of surface rotation
can be attributed to a very young age. The origin of activity in most of our CPM systems
lies in short-period binarity of their components, i.e., in hierarchical multiplicity.
An interesting connection emerges between the presence of wide companions and the
existence of short-period binaries. The reason for abundant multiple systems may partly
be purely dynamical, in that the chances of survival are higher for systems with an
internal binary because of the larger mass. An alternative astrophysical possibility
is that the original fragmentation of a star-forming core takes place at various spatial
scales and tends to produce multiple stellar systems, of which only hierarchical ones
can survive for an appreciably long time.

Apparently, the time-scale of dynamical survival of wide companions (of order 1 Gyr)
is sufficiently long compared to the time-scale of dynamical evolution of non-coplanar
multiple systems (the Kozai cycle, \S \ref{kozai.sec}) for the latter to shape up the
present-day systems. The existence of circularized spectroscopic binaries with periods
less than a few days may be the direct consequence of the interaction with remote companions,
followed by the tidal friction and loss of angular momentum \citep{eggl}. Ultimately,
the inner components will form a contact binary and then merge. The existence of
CPM multiple systems in a wide range of ages and separations will allow us to investigate
this process in detail as it unfolds. Indeed, even in our sample of modest size we
find examples of inner pairs of intermediate periods and large eccentricities, which are
apparently evolving toward the tidally circularized state. The Kozai-type mechanism can
affect the dynamical stability and composition of planetary systems. We find two stars
in our sample with multiple planets (55 Cnc and GJ 777 A), and both have interesting
dynamical properties very much unlike our Solar system.

\acknowledgments
The research described in this paper was in part carried out at the Jet Propulsion 
Laboratory, California Institute of Technology, under a contract with the National 
Aeronautics and Space Administration. This research has made use of the SIMBAD database,
operated at CDS, Strasbourg, France; and data products from the 2MASS, which is a joint project
of the University of Massachusetts and the Infrared Processing and Analysis Center, California
Technology Institute, funded by NASA and the NSF.

\clearpage
\begin{deluxetable}{ccccrrrrrrrrr}
\tabletypesize{\scriptsize}
\rotate
\tablecaption{Examined CPM double and multiple systems within 25 pc \label{big.tab}}
\tablewidth{0pt}
\tablehead{
\colhead{HIP/Name} & \colhead{Alt. name} & \colhead{RA2000} &
\colhead{DEC2000} & \colhead{Sep.} & \colhead{$\mu_\alpha\cos\delta$} & \colhead{$\mu_\delta$} &
\colhead{$\Pi[\sigma_\Pi]$} &\colhead{$V$ mag} & \colhead{$V-I$} & \colhead{$J$ mag} &
\colhead{$H$ mag} & \colhead{$K$ mag}}
\startdata
  473 & GJ 4 A & 00 05 41.0129 & +45 48 43.491 &       & 879 & -154 &  85.10[2.74] & 8.23 & 1.77 & 6.10 & 6.82 & 5.26 \\
  428 & GJ 2   & 00 05 10.8882 & +45 47 11.641 & 328.1 & 870 & -151 &  86.98[1.41] & 9.97 & 2.53 & 6.70 & 6.10 & 5.85 \\
\hline
   4849 & GJ 3071 AB & 01 02 24.5721 & +05 03 41.209 &       & 340 & 221 &  46.61[1.61] & 8.15 &   &6.20 & 5.68 & 5.51 \\
WD 0101+048 & GJ 1027 & 01 03 49.9093 & +05 04 30.840 & 1276.0 & 320 & 232 &     & 14.10 & 0.34 & 13.50 & 13.40 &13.42 \\
\hline
 4872 & GJ 49 & 01 02 38.8665 & +62 20 42.161 &       & 730 &  89 &  99.44[1.39] & 9.56 & 1.97 & 6.23 & 5.58 & 5.37 \\
V388 Cas & GJ 51 & 01 03 19.8653 & +62 21 55.930 & 294.7 & 732 &  80 &     & 13.78 & 3.32 & 8.61 & 8.01 & 7.72 \\
\hline
5799 & GJ 9045 A & 01 14 24.0398 & -07 55 22.173 &       & 124 & 278 &  41.01[0.89] & 5.14 & 0.54 & 4.40 & 4.02 & 4.06 \\
LTT 683 & GJ 9045 B & 01 14 22.4332 & -07 54 39.232 & 49.1  & 123 & 272 &     & 7.83 & 0.83 & 6.40 & 6.02 & 5.88 \\
\hline
9749 & GJ 9070 A & 02 05 23.6559 & -28 04 11.032 &      & 340 &  422 &  44.37[1.97] & 10.96 & 1.69 & 7.99 &  7.35 &  7.16\\
LTT 1097 & GJ 9070 B & 02 05 24.6587 & -28 03 14.570 & 58.0 & 324& 422 &    & 12.82 & 2.49 & 8.80 &  8.26 &  8.04 \\
\hline
14286 & GJ 3194 A & 03 04 09.6364 & +61 42 20.988 &       & 721 & -693 &  43.74[0.84] & 6.67 &     & 5.39 & 512 & 5.03 \\
LTT 1095 & GJ 3195 B & 03 04 43.4407 & +61 44 08.950 & 263.4 & 718 & -698 &      & 12.55 & 2.27 & 8.88 & 8.33 & 8.10 \\
\hline
14555 & GJ 1054 A & 03 07 55.7489 & -28 13 11.013 &      & -339 & -120 &  55.5[2.5] & 10.24 & 1.70 & 7.24 & 6.58 &  6.37 \\
LTT 1477 & GJ 1054 B & 03 07 53.3793 & -28 14 09.650 & 66.5 & -336 & -112 &     & 13.09 & 2.31 & 9.35&  8.78 &  8.52 \\
\hline
15330 & GJ 136 & 03 17 46.1635 & -62 34 31.160 &       &  1338 & 649 &  82.51[0.54] & 5.53 & 0.71 & 4.46& 4.04 &  3.99 \\ 
15371 & GJ 138 & 03 18 12.8189 & -62 30 22.907 & 309.2 &  1331 & 647 &  82.79[0.53] & 5.24 & 0.68 & 4.27&  3.87 &  3.86 \\ 
\hline
 17414 & GJ 9122 A & 03 43 52.5624 & +16 40 19.272 &        & 155 & -320 &  58.09[1.98] & 9.96 & 1.65 & 7.05 & 6.41 & 6.25 \\
 17405 & GJ 9122 B & 03 43 45.2490 & +16 40 02.138 & 106.5  & 159 & -313 &  61.40[2.37] & 10.81 & 1.94 & 7.53 & 6.91 & 6.69 \\
\hline
 21482 & V833 Tau & 04 36 48.2425 & +27 07 55.897 &       & 232 & -147 &  56.02[1.21] & 8.10 & 1.60 & 5.95 & 5.40 &  5.24 \\
WD 0433+270 & NLTT 13599 & 04 36 44.8902 & +27 09 51.594 & 124.0 & 226 & -153 &      & 15.81 & 0.80 & 14.60 & 14.23 & 14.14 \\
\hline
22498 & DP Cam & 04 50 25.0911 & +63 19 58.624 &       & 219 & -195 &  42.59[17.78] & 9.83 & 1.22 & 7.55 & 6.95 & 6.80 \\
G 247-35 &     & 04 50 21.6640 & +63 19 23.420 & 42.1  & 210 & -194 &      & 12.72 & 2.15 & 9.20 & 8.59 & 8.36 \\  %%5 new optical?
\hline
 25278 & GJ 202 & 05 24 25.4633 & +17 23 00.722 &        & 250 &  -7 & 68.2[0.9] & 5.00 &   & 4.43 & 4.03 & 4.04 \\
 25220 & GJ 201 & 05 23 38.3810 & +17 19 26.829 & 707.2  & 253 &  -5 & 69.8[1.5] & 7.88 & 1.25 & 5.88 & 5.38 & 5.23 \\
\hline
34065 & GJ 9223 A & 07 03 57.3176 & -43 36 28.939 &     & -103 & 389 &  61.54[1.05] & $5.27^b$ & 0.73 & 4.41 & 3.99 & 4.04 \\  
34069 & GJ 9223 B & 07 03 58.9171 & -43 36 40.857 & 21.1 &  -99 & 383 &  66.29[6.81] & 6.86 &   0.83 &  5.46 & 5.08 & 4.94 \\
34052 & GJ 264 & 07 03 50.24 & -43 33 40.7 & 184.8 & -93 & 395 &  58.89[0.94]  & $8.69^b$ & 1.35 & 6.45 & 5.83 & 5.70 \\
\hline
  42748 & GJ 319 A & 08 42 44.5315 & +09 33 24.114 &        & 216 &-634 &  74.95[13.82] &  9.63 & 1.86 & 6.69 & 6.05 & 5.83 \\
        & GJ 319 C & 08 42 52.2287 & +09 33 11.157 & 114.6  & 224 & -616 &    & 11.81 & 2.39 & 8.12 & 7.49 & 7.28 \\
\hline
 43587 & GJ 324 A & 08 52 35.8111 & +28 19 50.947 &       & -485 & -234 &  79.80[0.84] & 5.96 &      & 4.77 & 4.26 & 4.01 \\
LTT 12311 & GJ 324 B & 08 52 40.8393 & +28 18 59.310 & 84.1  & -488 & -234 &     & 13.14 & 3.00 & 8.56 &  7.93 & 7.67 \\
\hline
 46843 & GJ 9301 A & 09 32 43.7592 & +26 59 18.708 &       & -148 & -246 &  56.35[0.89] & 7.01 &    & 5.58 & 5.24 & 5.12 \\
NLTT 22015 & GJ 9301 B & 09 32 48.2450 & +26 59 43.864 & 65.0 & -142 & -243 &      &      &   & 10.36 & 9.86 & 9.47 \\
\hline
47620 & GJ 360 & 09 42 34.8429 & +70 02 01.989 &       & -671 & -269 &  85.14[3.18] & 10.58 & 2.20 & 6.92 & 6.33 & 6.08 \\
47650 & GJ 362 & 09 42 51.7315 & +70 02 21.892 & 88.8  & -669 & -264 &  86.69[2.24] & 11.24 & 2.41 & 7.33 & 6.73 & 6.47 \\
\hline
 49669 & GJ 9316 A & 10 08 22.3106 & +11 58 01.945 &        & -249 &   5 &  42.09[0.79] & 1.35 &    & 1.67 & 1.66 & 1.64 \\
       & GJ 9316 B & 10 08 12.7970 & +11 59 49.078 & 176.0  & -244 &  12 &     & 8.11 & 1.00 & 6.42 & 5.99 & 5.88 \\
%%%[photometry from CTIO run]
\hline
 50564 & GJ 9324 & 10 19 44.1679 & +19 28 15.290 &        & -230 & -215 &  47.24[0.82] & 4.80 &   & 4.04 & 3.94 & 4.02 \\
 NLTT 23781 &      & 10 14 53.8493 & +20 22 14.590 & 5220.6 & -232 &  -212 &     & 16.48 &   & 10.81 & 10.20 & 9.99 \\
\hline
59000 & GJ 9387 & 12 05 50.6574 & -18 52 30.916 &       & -19 & -320 &  44.41[1.51] & 9.95 & 1.57 & 7.42 & 6.79 &  6.62 \\
NLTT 29580 &    & 12 05 46.6407 & -18 49 32.240 & 187.6 & -4 & -314  &       & 16.23 & 3.32 & 11.20 & 10.63 & 10.32 \\
%%% [photometry from CTIO run]
\hline
59406 & GJ 3708 A & 12 11 11.7583 & -19 57 38.064  &      & -216 & -184 &  78.14 [2.80] & 11.68 & 2.33 & 7.89 &  7.36 &  7.04 \\
NLTT 29879 & GJ 3709 B & 12 11 16.95 & -19 58 21.9 & 85.2 & -203 & -188 &      & 12.62 & 2.51 & 8.60 &  8.01 &  7.74 \\ 
\hline
61451 & GJ 1161 A & 12 35 33.5525 & -34 52 54.901 &      & -228 & -134 & 46.19[0.91] & 7.87 & 1.07 & 5.95&  5.44 & 5.26 \\ 
LTT 4788 & GJ 1161 B & 12 35 37.7821 & -34 54 15.309 & 95.8 & -219 & -128 &   & 11.76 & 2.39  & 8.15 &  7.58 &  7.30 \\
%%%   [VI photometry from CTIO run]
\hline
 63882 & GJ 3760 & 13 05 29.8783 & +37 08 10.635 &       & -304 & -202 &  43.18[6.95] & 10.62 &     & 8.22 & 7.61 & 7.35 \\
 NLTT 33194 &    & 13 11 24.2045 & +37 24 37.197 & 4342.8 & -305 & -193 &           &  &     & 11.89 & 11.36 & 11.10 \\
%%%5 new
\hline
65083 & LTT 5136 & 13 20 24.9410 & -01 39 27.026 &      & 129 & -251 &  48.18[3.05] & 11.61 & 1.94 & 8.39 & 7.79 & 7.53 \\
LTT 5135 &       & 13 20 12.5595 & -01 40 40.980 & 199.8 & 129 & -257 &    & 13.41 & 2.39 & 9.57 & 8.98 & 8.78 \\
\hline
65877 & GJ 515 & 13 30 13.6398 & -08 34 29.492 &       & -1107 & -475 &  55.50[3.77] & 12.39 & -0.01 & 12.62& 12.68 &12.74 \\
LTT 5214 &    & 13 30 02.8247 & -08 42 25.530 & 502.3 & -1102 & -472 &      & 14.33 & 3.04 & 9.60 & 9.05 & 8.75 \\
\hline
 66492 & NLTT 34715 & 13 37 51.2257 & +48 08 17.079 &       & -234 & -139 &  45.66[2.72] & 9.77 & 1.60 & 6.94 & 6.34 & 6.14 \\
NLTT 34706 & GJ 520 C & 13 37 40.4407 & +48 07 54.169 & 110.4 & -225 & -136 &     & 14.47 & 2.73 & 10.12 & 9.59 & 9.30 \\
\hline
 71914 & GJ 9490 A & 14 42 33.6486 & +19 28 47.219 &        & -254 & -154 &  44.54[2.57] & 9.11$^b$ & 1.33 & 6.60 &  5.97 & 582 \\
 71904 & LTT 14350 & 14 42 26.2580 & +19 30 12.694 & 135.0  & -261 & -177 &  38.62[2.01] & 10.08 & 1.43 & 7.45 & 6.80 & 6.66 \\
\hline
75718 & GJ 586 A & 15 28 09.6114 & -09 20 53.050 &       & 73 & -363 &  50.34[1.11] & 6.89 & 0.87 & 5.44 & 5.05 & 4.89 \\
75722 & GJ 586 B & 15 28 12.2103 & -09 21 28.296 & 52.2  & 82 & -356 &  48.06[1.14] & 7.54 & 0.91 & 5.99 & 5.55 & 5.46 \\
\hline
 79607 & GJ 9550 A & 16 14 40.8536 & +33 51 31.006 &       & -266 & -87 &  46.11[0.98] & 5.70 & 0.80 & 3.95 & 3.35 & 4.05 \\
 79551 & GJ 9549   & 16 13 56.4533 & +33 46 25.030 & 632.3 & -264 & -84 &  43.82[5.69] & 12.31 & 3.06 & 8.60 & 8.00 &  7.75 \\
\hline
82817 & GJ 644 AB & 16 55 28.7549 & -08 20 10.838 &       & -829 & -879 &  174.22[3.90] & 9.02 & 2.33 & 5.27 & 4.78 & 4.40 \\
82809 & GJ 643 & 16 55 25.2251 & -08 19 21.274 & 72.1  & -813 & -895 &  153.96[4.04] & 11.74 & 2.63 & 7.55 & 7.06 & 6.72 \\
LHS 429 & GJ 644 C & 16 55 35.2673 & -08 23 40.840 & 231.2 & -810 & -872 & 154.5[0.7]$^c$ & 16.85$^c$ & 4.54$^c$ & 9.78 & 9.20 & 8.82 \\
\hline
83599 & GJ 653 & 17 05 13.7781 & -05 05 39.220 &      & -921 & -1128 &  89.70[28.71] & 10.09 & 2.13 & 6.78&  6.19 & 5.97 \\
83591 & GJ 654 & 17 05 03.3941 & -05 03 59.428 & 184.5 & -917 & -1138 &  92.98[1.04] & 7.73 & 1.35 & 5.52 & 4.94 & 4.73 \\
\hline
 86036 & 26 Dra & 17 34 59.5940 & +61 52 28.394 &       & 277 & -526 &  70.98[0.55] & 5.23 &     & 4.24 & 3.88 & 3.74 \\ %%% double
 86087 & GJ 685 & 17 35 34.4809 & +61 40 53.631 & 737.5 & 264 & -514 &  70.95[1.09] & 9.97 & 1.81 & 6.88 & 6.27 & 6.07 \\ 
\hline
 93899 & GJ 745 B & 19 07 13.2039 & +20 52 37.254 &       & -481 & -333 &  112.82[2.41] & 10.76 & 2.09 & 7.28 & 6.75 & 6.52 \\
 93873 & GJ 745 A & 19 07 05.5632 & +20 53 16.973 & 114.2 & -481 & -346 &  115.91[2.47] & 10.78 & 2.09 & 7.30 & 6.73 & 6.52 \\
\hline
 97295 & GJ 9670 A & 19 46 25.6001 & +33 43 39.351 &       & 19 & -446  &  47.94[0.54] & 4.96 & 0.53 & 4.05 & 3.98 & 3.83 \\
 97222 & LTT 15766 & 19 45 33.5520 & +33 36 06.055 & 792.3 & 23 & -449  &  49.09[1.43] & 7.68 &    & 5.81 & 5.32 & 5.25 \\ %%% res.double new
LTT 15775 & GJ 9670 B & 19 46 27.5446 & +33 43 48.894 & 25.8  & 25 & -438 &     & 8.58 & 1.13 & 6.64 & 6.12 & 6.00 \\
\hline
98204 & GJ 773 & 19 57 19.6421 & -12 34 04.746 &         & -94 & -513 & 52.92[1.48] & 9.31 & 1.36 & 6.82 & 6.20 & 6.02 \\
NLTT 48475 &   & 19 57 23.8000 & -12 33 50.260 & 62.6  & -92 & -518 &      & 15.36 & 3.39 & 10.21 & 9.65 & 9.32 \\
\hline
 98767 & GJ 777 A & 20 03 37.4055 & +29 53 48.499 &       & 683 & -524 &  62.92[0.62] & 5.73   &    &4.55 & 4.24 & 4.08 \\
 LTT 15865 & GJ 777 B & 20 03 26.5799 & +29 51 59.595 & 178.0 & 687 & -530 &      & 14.38 & 3.03 & 9.55 & 9.03 & 8.71 \\
\hline
102409 & GJ 803 & 20 45 09.5317 & -31 20 27.238 &        & 261 & -345 &  100.59[1.35] & 8.75 & 2.07 & 5.81 & 5.20 & 4.94 \\
102141 & GJ 799 & 20 41 51.1537 & -32 26 06.730 & 4680.0 & 280 & -360 &  97.80[4.65] & 10.33 & 2.92 & 5.44 & 4.83 &  4.53 \\
\hline
109084 & GJ 4254 & 22 05 51.2986 & -11 54 51.022 &       & -266 & -175 &  46.70[7.86] & 10.15 &      & 7.22 & 6.60 & 6.40 \\
LP 759-25 &      & 22 05 35.7280 & -11 04 28.820 & 3030.9 & -274 & -162 &         &   &      & 11.66 & 11.05 & 10.72 \\
\hline
113602 & NLTT 9310 & 23 00 33.4015 & -23 57 10.309 &      & 190 & -345 &  49.15[3.03] & 11.57 & 1.95 & 8.25 & 7.67 &  7.41 \\
113605 & NLTT 9315 & 23 00 36.5922 & -23 58 10.657 & 74.5 & 195 & -346 &  49.36[3.19] & 11.61 & 1.98 & 8.26 & 7.66 & 7.42 \\
\hline
115147 & V368 Cep & 23 19 26.6320 & +79 00 12.666 &       & 201 & 72 &  50.65[0.64] & 7.54 &      & 5.90 & 5.51 & 5.40 \\ %%%% parallax 50.7 \pm 0.6 B companion at 11" post-TT
LSPM J2322+7847 &  & 23 22 53.8733 & +78 47 38.810 & 959.1 & 210 & 64 &    & 16.18 & 3.62 & 10.42 & 9.84 & 9.52 \\
\hline
116215 & GJ 898 A & 23 32 49.3998 & -16 50 44.308 &       & 344 & -218 &  71.70[1.36] & 8.62 & 1.28 & 6.24 & 5.61 & 5.47 \\
116191 & GJ 897 & 23 32 46.5991 & -16 45 08.395 & 338.3 & 382 & -186 &  89.9[7.3] & 10.43 & 2.24 & 6.71 &  6.09 & 5.86 \\

\enddata
\tablecomments{Key
to the columns: (1) HIP number or name (2) alternative name; (3) Right Ascension (2000); 
(4) Declination (2000); (5) separation on the sky in arcsec; (6) and (7) proper motion in mas/yr; 
(8) parallax and its error in mas; (9) $V$ magnitude; 
(10) $V-I$ color; (11) $J$ magnitude; (12) $H$ magnitude;
(13) $K$ magnitude. Our photometric observations are marked with superscript $^b$, and
$^c$ denotes photometry and parallax from \citet{dahn}.}
\end{deluxetable}

\clearpage
\begin{deluxetable}{lrrrr}
\tabletypesize{\scriptsize}
\tablecaption{Vertical motion of Galactic components \label{z.tab}}
\tablewidth{0pt}
\tablehead{
\colhead{Component} & \colhead{$|v_{z0}|$} & \colhead{$z_{\rm max}$} & \colhead{$P_\nu$} & 
\colhead{$f(|z|<100)$}\\
\colhead{} & \colhead{\kms} & \colhead{pc} & \colhead{Myr} & \colhead{}\\
}
\startdata
  Thin disk (young) \dotfill & 6 & 72 & $79$ & $1.00$ \\
  Thin disk (giants)\dotfill & 18 & 217 & 86 & $0.31$ \\
  Thick disk        \dotfill & 35 & 422 & 109 & $0.16$ \\
  Halo              \dotfill & 94 & 1130 & 305 & $0.07$\\
\enddata
\end{deluxetable}

\clearpage
\begin{deluxetable}{rlrlrlrlrl}
\tabletypesize{\scriptsize}
\rotate
\tablecaption{WDS identifications and new pairs \label{wds.tab}}
\tablewidth{0pt}
\tablehead{
\colhead{HIP} & \colhead{WDS} & \colhead{HIP} & \colhead{WDS} & \colhead{HIP} & \colhead{WDS} & 
\colhead{HIP} & \colhead{WDS} &\colhead{HIP} & \colhead{WDS} }
\startdata
  473 & 00057+4549 &  4849 & new &  4872 & new &  5799 & new &  9749 & 02053-2803  \\
14286 & 03042+6142 & 14555 & 03079-2813 & 15371 & 03182-6230 & 17414 & 03439+1640 & 21482 & 04368+2708  \\
22498 & 04503+6320 & 25278 & new & 34065 & 07040-4337 & 42748 & 08427+0935 & 43587 & 08526+2820 \\
49669 & 10084+1158 & 46843 & 09327+2659 & 47620 & 09427+7004 & 50564 & new & 59000 & new  \\
59406 & 12113-1958 & 61451 & 12356-3453 & 63882 & 13055+3708 & 65083 & 13203-0140 & 65877 & 13303-0834 \\
66492 & 13379+4808 & 71914 & 14426+1929 & 75718 & 15282-0921 & 79607 & 16147+3352 & 82817 & 16555-0820 ($231"$ comp. is new) \\
83591 & 17050-0504 & 86036 & 17350+6153 & 93899 & 19072+2053 & 97295 & 19464+3344 ($792"$ comp. is new) & 98204 & 19573-1234 \\
98767 & 20036+2954 & 102409 & 20452-3120 & 109084 & new & 113602 & new & 115147 & 23194+7900 \\
116215 & 23328-1651 \\

\enddata
\end{deluxetable}

\begin{deluxetable}{rlllr}
\tabletypesize{\scriptsize}
%%\rotate
\tablecaption{X-ray luminosities \label{x.tab}}
\tablewidth{0pt}
\tablehead{
\colhead{HIP/Name} & \multicolumn{3}{c}{$L_X$} & \colhead{HR1} \\}
\startdata
15371 & $0.058$ & $\pm$ & $0.013$ & $-0.91$ \\ 
61451 & $0.11 $ & $\pm$ & $0.04$   & $-0.86$ \\ 
102409 & $5.59$ & $\pm$ & $0.11$ & $-0.07$ \\
102141 & $3.38$ & $\pm$ & $0.10$ & $-0.19$ \\
14555 & $4.62 $ & $\pm$ & $0.27$ & $-0.27$ \\
59000 & $0.71 $ & $\pm$ & $0.24$ & $+0.20$ \\
116215 & $0.13$ & $\pm$ & $0.03$ & $-0.56$ \\
116191 & $1.41$ & $\pm$ & $0.09$ & $-0.28$ \\
75722 & $0.43 $ & $\pm$ & $0.07$ & $-0.40$ \\
82817 & $1.09 $ & $\pm$ & $0.07$ & $-0.26$ \\
5799 & $0.74  $ & $\pm$ & $0.12$ & $-0.01$ \\
25278 & $0.16 $ & $\pm$ & $0.03$ & $-0.48$ \\
25220 & $2.75 $ & $\pm$ & $0.11$ & $-0.12$ \\
50564 & $1.09 $ & $\pm$ & $0.38$ & $+0.22$ \\
46843 & $1.49 $ & $\pm$ & $0.09$ & $-0.19$ \\
21482 & $8.20 $ & $\pm$ & $2.38$ & $-0.04$ \\
97222 & $0.087$ & $\pm$ & $0.023$ & $-0.85$ \\
LTT 15775 & $0.048$ & $\pm$ & $0.06$ & $-0.29$ \\
79607 & $46.1 $ & $\pm$ & $0.6$ & $+0.06$ \\
  473 & $0.050$ & $\pm$ & $0.014$ & $-0.42$ \\
 66492 & $0.076$& $\pm$ & $0.028$ & $-0.74$ \\
86036 & $0.47 $ & $\pm$ & $0.02$ & $-0.48$ \\
86087 & $0.028$ &$\pm$ & $0.005$ & $-0.58$ \\
V388 Cas & $0.20$&$\pm$ & $0.02$ & $-0.19$ \\
47650 & $0.18 $ & $\pm$ & $0.03$ & $-0.40$ \\
115147 & $10.6$ & $\pm$ & $0.2$ & $-0.10$ \\
\enddata
\tablecomments{Key
to the columns: (1) HIP number or name (2) X-ray luminosity in units of $10^{29}$
erg s$^{-1}$ (3) hardness ratio $HR1$ from ROSAT.}
\end{deluxetable}

\clearpage
\begin{deluxetable}{llrrrrr}
\tabletypesize{\scriptsize}
\tablecaption{Velocities, moving groups, activity and ages \label{kin.tab}}
\tablewidth{0pt}
\tablehead{
\colhead{HIP/Name} & \colhead{$U$} & \colhead{$V$} & \colhead{$W$} & \colhead{SKG(ref)} &
\colhead{EW(H$\alpha$)} & \colhead{$\log$Age}\\}
\startdata
15330 & $-71$ & $-47$ & $+16$ & Hercules ? (1) &   & $9.2$ \\ 
15371 & $-70$ & $-46$ & $+16$ & Hercules ? (1) &   & $9.4$ \\ 
47620 & $-36$ & $-13$ & $-14$ & Hyades     (2) & $<0$ & $>8.7$ \\
47650 & $-35$ & $-13$ & $-14$ & Hyades     (2) & 2.87 & $<8.9$ \\ 
79607 & $-7 $ & $-29$ & $+9 $ &                & 0.64 & $>9.5$ \\
21482 & $-39$ & $-17$ & $-2 $ & Hyades?    (3) & 1.0  & $>9.6$ \\
102409& $-10$ & $-17$ & $-10$ & BETAPIC    (4) & 2.2  & $7.0 $ \\
102141& $-9 $ & $-16$ & $-11$ & BETAPIC    (4) & 10.9 & $7.0 $ \\
25278 & $-37$ & $-15$ & $8  $ & Hyades     (2) & $  $ & $8.5 $ \\
25220 & $-38$ & $-14$ & $7  $ & Hyades     (2) & $-0.76$& $8.7 $ \\ 
116215 & $-13$ & $-21$ & $-10$&                & $-0.59$& $8.8 $ \\ 
116191 & $-13$ & $-21$ & $-10$&                & $1.98$& $<9.0 $ \\ 
4872 & $-32$ & $-16$ & $6  $  & Hyades     (2) & $  $ & $ $ \\
43587 & $-37$ & $-18$ & $-8$  & Hyades     (2) & $  $ & $ $ \\

\enddata
\tablecomments{References: (1) \citet{soub}; (2) \citet{mont}; (3) \citet{egg93}; (4) \citet{ma07}.}
\end{deluxetable}

\end{document}